\documentstyle[12pt]{article}
\begin{document}
\begin{titlepage}
\vspace*{16mm}
\begin{center}
{\Large
{\bf
    Momentum Lattice Simulation on a Small Lattice Using Stochastic
    Quantization
}}
\end{center}
\vspace{5mm}
\begin{center}
H. Kr{\"o}ger$^1$, S. Lantagne$^1$ and K.J.M. Moriarty$^2$\\
\end{center}
{\small $^1$ \space D{\'e}partement de Physique, Universit{\'e} Laval,
              Qu{\'e}bec, Qu{\'e}bec, G1K 7P4, Canada \\
        $^2$ \space Department of Mathematics, Statistics and Computing
                    Science, Dalhousie University,  \\
             \hspace*{4mm} Halifax, N.S., B3H 3J5, Canada}
\begin{abstract}
   We have studied the scalar $\phi^4$-model in the symmetric phase and the
 non--compact $U(1)$ gauge theory on
a momentum lattice using the Langevin equation for generating configurations.
In the  $\phi^4$-model we have analyzed the renormalized mass and
in the $U(1)$-model
 we have analyzed
the Wilson loop operator. We
used a second order algorithm for solving the Langevin equation, and we
 looked for the convergence rate of the method. We studied  the stochastic
time  needed
to generate equilibrium configurations and compared first and second order
schemes for both models.

\end{abstract}
\end{titlepage}
\section{Introduction}

    One major problem in lattice gauge theory has been the increased amount
of computer time and memory needed  to do simulations near a critical point.
 The so-called critical slowing
down can be countered, e.g., with the Fourier acceleration method \cite{Batr1}.
 Many other methods have been
developed to accelerate   convergence of the calculation, e.g., the cluster
algorithm \cite{Swendsen}, suitable for Ising-type systems.
When   we
 studied the scalar $\phi^4$-model near the critical point
on a {\it momentum lattice} \cite{Lant}, it turned out that quite reasonable
results for the renomalized
mass, wave function renormalization, etc., could be obtained on relatively
small lattices ($3^4$---$7^4$). In this note we want to report a more detailed
numerical analysis of the convergence behavior on a momentum lattice.
\section{Momentum Space}

    Momentum space has several interesting advantages over  coordinate space:
\begin{itemize}
   \item The kinetic energy  of the action is local.
   \item One can implement Fourier acceleration to fight critical slowing down.
   \item At the critical point, the correlation length goes to infinity.
         The behavior at $x \rightarrow \infty$ in coordinate space
         corresponds
         to $k \rightarrow 0$ in momentum space.
   \item The correlation function in momentum space behaves as
         $1 / ( m^2_R + k^2 )$ near the critical point. At the critical point,
         $m_R \rightarrow 0$, but one can stay away from the pole by choosing
         some $k^2 \neq 0$.
   \item Although the action in momentum space is non-local, one can use a fast
         Fourier transform to switch to coordinate space, where the interaction
         is local.
\end{itemize}
\section{Stochastic Quantization}

    We have used stochastic quantization, which is a direct
method to go from the
continuum formulation to practical algorithms for numerical simulations. We
have studied the $\phi^4_{3+1}$ model and the non-compact $U(1)_{3+1}$ gauge
 model. Stochastic quantization was introduced by Parisi and Wu
 \cite{ParisiWu}, and
it has been applied by many authors (for a review see Damgaard and H{\"u}ffel
 \cite{DHColl,DamHuf}). The
idea is to consider the Euclidean quantum field as the equilibrium state
of a statistical system coupled to a heat reservoir. The evolution of this
 statistical
system is described by the Langevin equation, which reads in momentum space,
\begin{displaymath}
    \frac{\partial \hat \phi(k,\tau)}{\partial \tau} =
    - \frac { \delta S[\hat \phi]} {\delta \hat \phi(-k,\tau)}
    + \hat \eta(k,\tau),
\end{displaymath}
where $S[\hat \phi]$ is the action in momentum space and $\hat \eta(k,\tau)$ is
a field of Gaussian noise. When $\tau \rightarrow \infty$  the statistical
system is ``equal'' to the Euclidean field; $\tau$ was
introduced as fictitious time to evolve the system. Our goal is to solve the
Langevin equation many times to generate a good representation
of the canonical ensemble. Then we perform the canonical
 averaging of the desired observable.
\section{Numerical Point of View}

    The expectation value of an observable is defined by
\begin{displaymath}
   {\overline {O[\hat \phi]}} = \lim_{\tau \rightarrow \infty} \frac{1}{\tau}
  \int_{0}^{\tau}
  \!\! d \tau' \, O[\hat \phi(\tau')].
\end{displaymath}
Because we suppose that our system is ergodic, the expectation value can
be computed over the configuration space,
\begin{displaymath}
  \left < O[\hat \phi] \right > = \lim_{N \rightarrow \infty } \frac{1}{N}
   \sum^N_{n=1}
   O[\hat \phi_n].
\end{displaymath}
So one must solve the Langevin equation $N$
times in parallel, which can be done using a parallel processor.

    The Langevin equation is a first order differential equation that
we integrate up to a final value  $\tau_{final}$, by increments of
$\Delta \tau$. It can
be solved using a standard method like Euler's method where the truncation
term is $O(\Delta \tau^2)$, or a higher order scheme like Heun's method
 (second order
Runge-Kutta) where the truncation term is $O(\Delta \tau^3)$.
Batrouni et al. \cite{Batr1} have shown that a higher order integration scheme
 has many
advantages, and our experience \cite{Lant} agrees with theirs.
For the same level of error, Heun's method is faster; it takes fewer
``time'' steps. For the same ``time'' step, it was 46\% slower for the
$\phi^4$-model,
but in the case of the non-compact $U(1)$-model it was only 12\% slower;
in the case of the $\phi^4$-model, the second order scheme required
the evaluation of two
additional fast Fourier transforms, but in the case of the $U(1)$-model
 only a few more additions
and multiplications were required. In Fig.[1] we display for the $\phi^4$-model
 the behavior of the
renomalized mass $m_R$  in the neighborhood of
the critical point. We have
computed in Fig.[2] the renormalized mass $m_R$  as a function of
$\tau_{final}$ for  large $\Delta \tau$ on a lattice of $3^4$ points.
This corresponds to
a coupling parameter $\kappa = 0.12025$, while the critical point is
 at $\kappa_c = 0.1257(1)$.
 Stability of the solution is obtained for a  value of
 $\tau_{final} \approx 12.5$ by Heun's method. For
larger values of $\tau_{final}$, the accumulation of error becomes
important and
fluctuations can appear. We can see that the value of $m_R$ drops beyond
$\tau_{final} \approx 25$
 in Fig.[2]. It is easier to determine the stability region in the second
order method.

   For the non-compact $U(1)$-model we have evaluated the expectation value
\linebreak
$\left < \sum_{\mu, \nu = 1 }^{4}  F_{\mu \nu}F_{\mu \nu} \right >$,
 where $F_{\mu \nu}$ is the field tensor.
 We get very good agreement with the known exact value;
an error of less than 1\% can be obtained easily. In momentum space
 $\left < F_{\mu \nu}F_{\mu \nu} \right >$
 is a local observable.

We also computed the expectation
value $W(I,J)$ of the Wilson loop observable, which corresponds to the
creation of a
quark-antiquark pair,
the propagation of this ``meson'' and the destruction of
the pair ($I \times J$ being the size of the rectangular loop measured in
units of the lattice spacing).
It is a non-local observable that can be computed analytically in
the $U(1)$-model. It seems
to be more sensitive to fluctuations than $\left < F_{\mu \nu}F_{\mu \nu}
\right >$.
 When
 integrating the Langevin equation, if we continue to iterate too long after
 the
equilibrium is reached, the field of noise $\hat \eta(k,\tau)$ creates some
fluctations in the solution. These fluctuations are more important for
non-local observables, which are difficult to evaluate on small lattices.
In Figs.
[3] and [4], we computed the expectation value of the Wilson loop operator for
different sizes of loops and for two different ``time'' steps on a lattice of
$6^4$ points. We
can see the fluctuation introduced by the noise field $\hat \eta(k,\tau)$. In
Fig.[3] the ``time'' step was $\Delta \tau = \tau_{final}/100$, and in
Fig.[4] the ``time'' step was $\Delta \tau = \tau_{final}/150$. The
important difference is the number of ``time'' steps.
When we compare the results of Figs. [3,4] we notice a difference which seems
to exceed the error ${\cal O}(\Delta \tau^2)$ due to finite $\Delta \tau$.
 We can reduce fluctuations by increasing the number of
configurations of
fields, but we can also average  several values of $\tau$ up
to $\tau_{final}$. Although we have not used the last method, we think it
 might reduce fluctuations \cite{Creutz}.

     Stochastic quantization allows us  to choose the value of
$\tau_{final}$. The propagator for the free scalar theory in Euclidean momentum
space for finite $\tau_{final}$ is
\begin{displaymath}
   \left < \hat \phi(k) \hat \phi(k') \right > = \delta(k+k')
   \frac{1-\exp(-2\tau_{final}
     (k^2+m^2))} {k^2+m^2},
\end{displaymath}
 compared with the Euclidean Feynman propagator,
\begin{displaymath}
    \left < \hat \phi(k) \hat \phi(k') \right > = \delta(k+k')
    \frac{1}{k^2+m^2}.
\end{displaymath}
We can see that one must choose $2\tau_{final}(k^2+m^2))$ large enough to make
the exponential term negligible. For the non-compact $U(1)$-model
in momentum space, we are free to choose a value for each lattice point
(Fourier
acceleration).
The point $k = 0$ turns out to be dangerous in the $U(1)$-model because there
 is no
interaction between lattice sites. For the $\phi^4$-model that does not seem
 to be
the case; the convergence rate seems to be a little slower near the critical
point, but does not differ very much.
\section{Conclusion}

    Our study of two models shows the feasibility of doing lattice
simulations on a small momentum lattice. The second order scheme was by far
more appropriate because of its greater accuracy and speed. The difference
in the results due to variation of $\Delta \tau$
encountered during the simulation of the $U(1)$-model for the
Wilson loop operator does not seem to pose a major problem.
We believe that the averaging over several values of $\tau$ up to
$\tau_{final}$ can increase the numerical stability.
This kind of behavior has not been seen in the $\phi^4$-model.
  The knowledge
of the stability region is very important for simulations of theories like
$SU(2)$ and $SU(3)$. For these models, the time needed for generating
equilibrium configurations is much larger than for the $U(1)$-model.
\vspace*{10mm}

{\large \bf Acknowledgments:}
\vspace*{3mm}

  We would like to thank Ken Streach and Terence R.B. Donahoe of the
Government of Nova Scotia for their continued interest, support, and
encouragement and grant support; HNSX Supercomputer, Inc., for HNSX
Supercomputer Fellowship Awards (Grants 90HSFA01, 90HSFA02, 90HSFA03,
and 90HSFA04); James R. Berrett, Samuel W. Adams and Paul W. Crum for their
continued interest, encouragement and support; The Natural Sciences and
Engineering Research Council of Canada (Grant Nos. NSERC A8420 and
NSERC A9030) for financial support; the Atlantic Canada Opportunities
Agency (AAP Project No. 2060-392,324) for financial support; The Nova
Scotia Business Development Corporation (Grant 92NSBDC01) for financial
support; the National Research Council of Canada (Grants 92/01517E and
92/02158E) for financial support; and the N.S. Department of Economic
Development (Grants SBCEP-306-92-155 and 92ATDF01) for further
financial support. H.K. wishes to acknowledge support by FCAR Qu{\'e}bec
and NSERC Canada.
\pagebreak
\newpage
\begin{flushleft}
{\bf Figure Captions}
\end{flushleft}
\begin{description}
\item[{Fig.1}]
Renormalized mass $m_R$ of the $\phi^4$-model versus coupling parameter
$\kappa$. The semi-analytical results are taken
from  L{\"u}scher and Weisz Ref.~\cite{Luscher}. The full curve is a fit to
 L{\"u}scher and Weisz's
data to guide the eye.
\item[{Fig.2}]
Renormalized mass $m_R$ of the $\phi^4$-model versus stochastic time
$\tau_{final}.$
\item[{Fig.3}]
Wilson loop $W(I,J)$ of the $U(1)$-model versus coupling constant $g$;
$\Delta \tau
= \tau_{final}/100$.
\item[{Fig.4}]
Same as Fig.[3] but  $\Delta \tau = \tau_{final}/150.$
\end{description}
\end{document}